\documentclass[aps,twocolumn,prb,nopacs,floatfix,nofootinbib]{revtex4-1}

\usepackage[T1]{fontenc}
\usepackage{helvet}
\usepackage{lmodern}
\usepackage{amsmath}
\usepackage{amssymb}
\usepackage{graphicx}
\usepackage{multirow}
\usepackage{amsfonts}
\usepackage{rotating}
\usepackage{placeins}
\usepackage{booktabs}
\usepackage{cancel}
\usepackage{color}
\providecommand{\U}[1]{\protect\rule{.1in}{.1in}}
\newcommand{\Hb}{$J_1$-$J_2$-$J_3$-$K$}
\begin{document}

\title{Effect of magnetic frustration on nematicity and
  superconductivity in Fe chalcogenides}

\author{J.~K.~Glasbrenner,$^{1*}$  I.~I.~Mazin,$^{2}$ Harald~O.~Jeschke,$^{3}$  P.~J.~Hirschfeld,$^{4}$ and Roser~Valent\'\i$^{3}$}
\affiliation{$^{1}$National Research Council/Code 6393, Naval Research Laboratory, Washington, DC 20375, USA}
\affiliation{$^{2}$Code 6393, Naval Research Laboratory, Washington, DC 20375, USA}
\affiliation{$^{3}$Institut f\"ur Theoretische Physik, Goethe-Universit\"at Frankfurt, 60438 Frankfurt am Main, Germany}
\affiliation{$^{4}$University of Florida, Gainsville, FL, USA}
\affiliation{$^{*}$e-mail: james.glasbrenner.ctr@nrl.navy.mil}

\date{\today}

\begin{abstract}
  Over the past few years Fe chalcogenides (FeSe/Te) have advanced to
  the forefront of Fe-based superconductors (FeBS) research. The most
  intriguing results thus far are for intercalated and monolayer FeSe,
  however experimental studies are still inconclusive.  Yet, bulk FeSe
  itself remains an unusual case when compared with pnictogen-based
  FeBS, and may hold clues to understanding the more exotic
  FeSe-derivatives.  The FeSe phase diagram is unlike the pnictides:
  the orthorhombic distortion, which is likely to be of a
  ``spin-nematic'' nature in numerous pnictides, is not accompanied by
  magnetic order in FeSe, and the superconducting transition
  temperature $T_{c}$ rises significantly with pressure before
  decreasing. In this paper we show that the magnetic interactions in
  chalcogenides, as opposed to pnictides, demonstrate unusual (and
  unanticipated) frustration, which suppresses magnetic, but not
  nematic order, favors ferro-orbital order in the nematic phase and
  can naturally explain the nonmonotonic pressure dependence of the
  superconducting critical temperature $T_{c}(P)$.
\end{abstract}

\pacs{Pacs Numbers: }
\maketitle

While full consensus regarding the mechanism of high-temperature
superconductivity in Fe-based superconductors (FeBS) remains elusive,
nearly all researchers agree that it is unconventional and that it has
a magnetic origin
\cite{Chubukov2012_ARCMP,Hirschfeld2011_RPP}. However, there is a
divergence of opinions on the nature of the electrons responsible for
magnetism. There is an itinerant approach based on calculating the
spin susceptibility with moderate Coulomb (Hubbard) and Hund's
interactions
\cite{Wang2009_EPL,Chubukov2008_PRB,Stanev2008_PRB,Kuroki2008_PRL,
  Graser2009_NJP,Sknepnek2009_PRB,Yao2009_NJP, Guterding2015_PRB} as
well as a localized approach where itinerant electrons responsible for
conduction and the Fermi surface interact with local spins
\cite{Seo2008_PRL,Lv2010_PRB}. Finally, there is an increasingly
popular description where the electrons have a dual character and
provide the local moments, the interaction between them, and the
electronic
conductivity~\cite{Dai2012_NatPhys,Moon2010_PRB,Lee2010_PRB,Yin2011_NatMater}.
Within this picture, FeBS can still be reasonably mapped onto a
short-range model of pairwise interactions between the local moments.

Following the discovery of the FeBS, there were multiple attempts to
map the exchange interactions onto the Heisenberg model. The
$J_{1}$-$J_{2}$ model on the square lattice~\cite{Chandra1990_PRL}
with nearest- ($J_1$) and next-nearest-neighbor ($J_2$) exchange
couplings was a natural starting point
\cite{Xu2008_PRB,Fang2008_PRB,Ma2008_PRB,Si2008_PRL}, but required
dramatically different couplings for ferro- and antiferromagnetic
neighbors, $J_{1a}\ll J_{1b}$ to reproduce the observed spin
waves~\cite{Zhao2009_NatPhys,Diallo2009_PRL} and \textit{ab-initio}
calculations~\cite{Yaresko2009_PRB}; it also failed to describe the
double-stripe configuration in FeTe~\cite{Li2009_PRB,Bao2009_PRL}.
The model was extended to include third-neighbor exchange $J_{3}$
\cite{Ma2009_PRL} to reproduce the FeTe magnetic ground
state. However, only the Ising model has this configuration as a
solution, and in the Heisenberg model it is not a ground state for any
set of parameters \cite{Ferrer1993_PRB,Sindzingre2010_JPCS}.
Therefore adding $J_{3}$ does not solve the problem. Besides, the
$J_{1a}\ll J_{1b}$ implies an unphysical temperature dependence of the
exchange constants (as $T$ approaches $T_{N}$, by symmetry
$J_{1a} \rightarrow J_{1b}$).

There were attempts to overcome these problems by adding the
nearest-neighbor biquadratic exchange interaction
$K(\mathbf{S}_{i}\cdot \mathbf{S}_{j})^{2}$ to the $J_{1}$-$J_{2}$
\cite{Yaresko2009_PRB,Wysocki2011_NatPhys,Wysocki2013_JPCS} or
$J_{1}$-$J_{2}$-$J_3$~\cite{Hu2012_PRB} Heisenberg model.  The
three-neighbor Heisenberg model with biquadratic term (denoted {\Hb}
model from now on) eliminates the need for the $J_{1a,1b}$ anisotropy
of the nearest-neighbor exchange and, for sufficiently large $K$ and
$J_{3}$, has a ground state consistent with that of FeTe.  The
biquadratic coupling in this model is also essential to explain the
splitting between the antiferromagnetic and orthorhombic phase
transitions in the Fe pnictides
\cite{Fernandes2012_PRB,Xu2008_PRB,Mazin2009_PhysicaC}.

Whereas the magnetism in Fe-pnictides is successfully explained by the
{\Hb} model, the Fe-chalcogenides remain problematic.  Specifically,
there are two important unresolved controversies regarding bulk FeSe;
(i) it shows a structural transition at $T_s \sim 90$~K but, contrary
to the Fe-pnictides, no magnetic order is observed below $T_s$.
Instead, an extended nematic region is detected
\cite{Baek2014_NatMat,Bohmer2015_PRL} and the system becomes
superconducting at $T_c \sim 8$ K.  (ii) The superconducting $T_{c}$
first increases with pressure and then decreases, forming a dome
\cite{Medvedev2009_NatMat}. This is in apparent contradiction with the
expectation of a decreasing $T_c$ with pressure when magnetism is
absent.

In the present work we propose a solution to this mystery and
generalize the results to the family of Fe-chalcogenides FeSe/Te. We
show, using {\it ab-initio} density functional theory calculations and
effective model considerations, that many properties of FeSe/Te are
related to its unusual magnetic frustration, absent in the
Fe-pnictides.  We show that {\Hb} is the minimal spin model that
includes the relevant complexity of the magnetism in Fe-chalcogenides.
We then identify a new range of parameters appropriate for FeSe/Te
where a highly competitive novel ``staggered dimer'' phase
\cite{Landau1985_PRB} is stabilized (recently shown to be the ground
state of FeSe in \textit{ab-initio} calculations
\cite{Cao2014_arXiv}).

\vspace{0.2cm}

\textbf{Exchange model and phase diagram}

We define the {\Hb} model on the square lattice as
\begin{eqnarray}
  H &=&\sum_{\text{nn}}\left[ J_{1}\mathbf{\hat{m}}_{i}\cdot \mathbf{\hat{m}}%
        _{j}+K\left( \mathbf{\hat{m}}_{i}\cdot \mathbf{\hat{m}}_{j}\right) ^{2}%
        \right]   \label{eq:1} \\
    &+&\sum_{\text{2nn}}J_{2}\mathbf{\hat{m}}_{i}\cdot \mathbf{\hat{m}}%
        _{j}+\sum_{\text{3nn}}J_{3}\mathbf{\hat{m}}_{i}\cdot \mathbf{\hat{m}}_{j}
        \notag
\end{eqnarray}
The first sum is taken over all nearest neighbor $\{i,j\}$ pairs of Fe
spins, the second one over all next nearest neighbors, {\it etc}.
$\mathbf{\hat{m}}$ is the unit vector in the spin direction,
$|\mathbf{\hat{m}}_{i}|\equiv 1$.  For $K=0(\infty )$, this model
reduces to the already solved Heisenberg \cite{Sindzingre2010_JPCS}
(Ising\cite{Landau1985_PRB}) on the square lattice.

\begin{figure*}[t]
\centering
\includegraphics[width=0.48\textwidth]{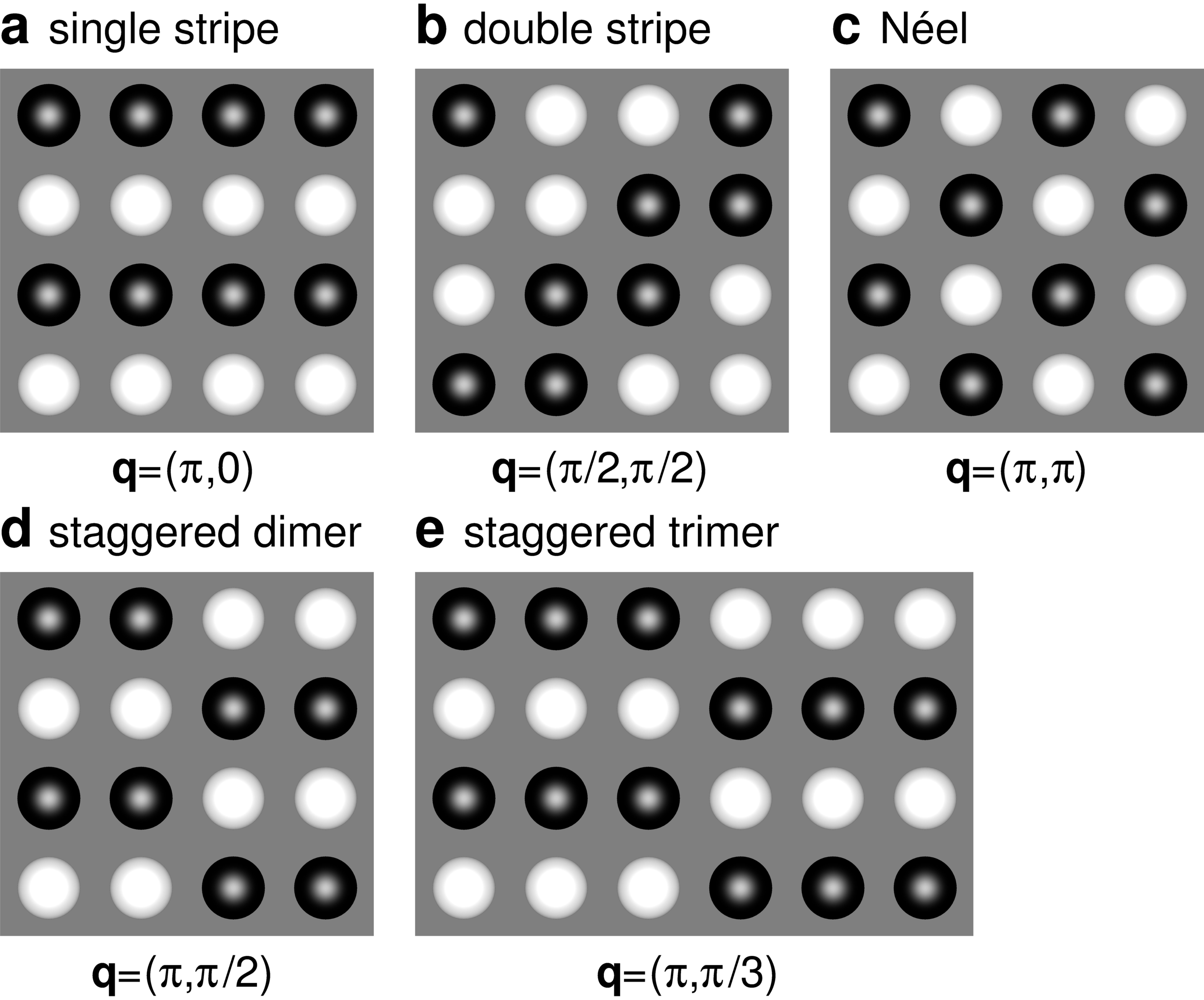}
\caption{\textbf{Collinear magnetic structures used for fitting to the
    {\Hb} model.} (a) single stripe, (b) double stripe, (c)
  checkerboard (N{\'e}el), (d) staggered dimer, and (e) staggered
  trimer state.}
\label{fig:magstructures}
\end{figure*}

To begin, we review the phase diagrams for the standard
$J_{1}$-$J_{2}$-$J_{3}$ Ising and Heisenberg models. In
Fig.~\ref{fig:phasediagram}a we show the mean-field $T=0$ Ising phase
diagram which includes the staggered dimer and double stripe ground
states (see Fig.~\ref{fig:magstructures} for pattern
definition). While this phase diagram can explain the
\textit{ab-initio} magnetic states of FeSe and FeTe, note that the
Ising model is inapplicable to low-anisotropy materials such as the
pnictides and chalcogenides, and the Heisenberg model is more
appropriate~\cite{Yaresko2009_PRB,Wysocki2011_NatPhys,Wysocki2013_JPCS,Hu2012_PRB}.
The mean-field phase diagram for the Heisenberg model at $T=0$ is
shown in Fig.~\ref{fig:phasediagram}b (quantum corrections introduce
minor changes \cite{Sindzingre2010_JPCS}).  The double stripe phase
has measure zero [it is a degenerate case of $\mathbf{q}=(Q,Q) $].
This means that \textit{no Heisenberg model can explain the formation
  of a collinear double stripe state}.

The review of the Ising and Heisenberg phase diagrams elucidates the
two theoretical problems that have been underemphasized in previous
analyses of the magnetic interactions of the Fe-based superconductors,
especially in the chalcogenides: (1) the Heisenberg model does not
account for all relevant magnetically ordered states and, by
implication, does not properly describe spin fluctuations, and (2) the
single and double stripe magnetic states are not the only important
ground state candidates for the chalcogenides; there is a third one,
the staggered dimers, which is highly competitive, but has been
routinely ignored. To address these problems the biquadratic term,
$K$, needs to be quantitatively taken into
account.

We solved the full {\Hb} model (Equation \eqref{eq:1}) for general $K$
in the mean-field limit and found six possible ground states (see
Supplementary Materials). Hu \textit{et.~al.}~\cite{Hu2012_PRB}
attempted prevously to solve this model, but missed the staggered
dimer phase \cite{Cao2014_arXiv} which, we argue, is the key to
understanding FeSe. A representative example phase diagram with
$K=0.1$ is shown in Fig.~\ref{fig:phasediagram}c. For a small, but
non-zero $K$ and $J_{3}$ the staggered dimer phase becomes stable in a
narrow ($|J_{2}-J_{1}/2|<2\sqrt{2KJ_{3}})$ interval near the critical
value $J_{1}=2J_{2}$, and at sufficiently large $J_{3}$
$(J_{3}>J_{1}^{2}/8K)$ the collinear double stripe structure is
stabilized. As $K$ grows, these collinear regions also grow, and at
$K>J_{1}/2$ the phase diagram becomes identical to the Ising phase
diagram in Fig.~\ref{fig:phasediagram}a. Actual materials will be seen
to lie in the intermediate region, $0<K\lesssim J_{1}/2$.

\vspace{0.2cm}

\textbf{First-principles calculations}

We performed density functional theory (DFT) calculations to obtain
parameters for Equation \eqref{eq:1} and place FeSe and FeTe into the
context of the {\Hb} phase diagram. There is a caveat though: due to
the itinerant character of magnetism in FeBS, mapping onto local
moments models such as Equation \eqref{eq:1} has limited accuracy. A
fundamental assumption of the standard Heisenberg model is that the
magnetic moments are rigid, and this is an excellent assumption for
systems with highly localized electrons, such as the high-$T_{c}$
cuprates, but relatively poor for itinerant electrons. Magnetic
interactions in metals tend to have long range tails, non-pairwise
interactions, and the moments may depend on the magnetic ordering
pattern. A clear example of the failure of the Heisenberg-biquadratic
models is that the double stripe (Fig.~\ref{fig:magstructures}b) and
plaquette (see Fig.~1p in the Supplementary Material) configurations
are degenerate in any such model, but in DFT the double stripe is 8
meV/Fe lower in energy than the plaquette configuration
\cite{Glasbrenner2014b_PRB}. Therefore, we cannot expect to derive a
Heisenberg model, with or without the biquadratic $K$, that captures
exactly the energetics of all possible magnetic configurations.

Despite these limitations, the {\Hb} model is the simplest framework
that accounts for all the magnetic ground states that DFT and
experiment find in different FeBS, and arguably is also the most
complex one that still allows for an analytic solution. Since we are
interested in spin fluctuation-driven effects such as
superconductivity and spin-nematicity, which are low-energy phenomena,
we establish a set of criteria for our fits, with the main goal to
select a consistent set of magnetic states and obtain parameters that
reproduce the low-energy hierarchy obtained within DFT. The criteria
are detailed in the Methods section, and the chosen magnetic
structures are shown in Fig.~\ref{fig:magstructures} panels a through
e.

We performed calculations for FeSe at three representative pressures
of 0, 4, and 9 GPa, and for FeTe at ambient pressure, see the Methods
and Supplementary Materials for details. In all cases we used
experimental lattice and internal parameters in tetragonal structures,
as discussed in Methods. We fitted to the five magnetic configurations
reported in Fig.~\ref{fig:energies} and extracted the $J_{1}$,
$J_{2}$, and $J_{3}$ parameters. The biquadratic term was extracted
from noncollinear calculations as in
Ref.~\onlinecite{Glasbrenner2014b_PRB}. The resulting {\Hb} model
parameters are reported in Table \ref{tab:exchangeparameters}.  Note
that the error bars reflect the fit inaccuracy, and not the much
smaller errors of the underlying DFT calculations.

First of all, we confirmed that the {\textquotedblleft}staggered
dimer{\textquotedblright} configuration \cite{Landau1985_PRB} is 13
meV/Fe lower in energy than the single stripe configuration and is the
true DFT ground state for FeSe \cite{Cao2014_arXiv} (see
Fig.~\ref{fig:energies}). The same phase is also the lowest in energy
in FeTe, as long as one does not take into account the magnetoelastic
coupling.  The calculated energy difference between the double stripe
and staggered dimer configurations in tetragonal FeTe is tiny,
$\sim 1-2$ meV/Fe. However, upon full structural relaxation into a
monoclinic structure the double stripe pattern gains more
magnetoelastic energy than the staggered dimer one (which relaxes into
an orthorhombic structure) and ends up lower by a few meV, with the
crystallographic distortion in agreement with
experiment~\cite{Li2009_PRB,Bao2009_PRL}.

Another important result is that while the main contenders for the
ground state of FeTe are the double stripe
($\mathbf{q}_{\mathrm{ds}}=(\pi /2,\pi /2)$) and the staggered dimer
($\mathbf{q}_{\mathrm{di}}=(\pi ,\pi /2)$) structures, with the
staggered trimers ($\mathbf{q}_{\mathrm{tri}}=(\pi ,\pi /3)$) a close
third, in FeSe the double stripe structure is not competitive at all.
In FeSe the lowest energy states are the staggered dimers, trimers,
tetramers and single stripes, with $\mathbf{q}$, respectively,
$(\pi ,\pi /2)$, $(\pi ,\pi /3)$, $(\pi ,\pi /4)$, and
($\mathbf{q}_{\mathrm{ss}}=(\pi ,0)$). From this, one can conclude
that while in experiment the long range order of FeSe is destroyed by
spin fluctuations, the most relevant ones are those with the
corresponding wave vectors as listed above, and, very likely, with any
$\mathbf{q}=(\pi ,Q)$ such that $0\leq Q\leq \pi /2$.

Importantly, when FeSe is structurally optimized in any of the low
energy magnetic structures, it admits an orthorhombic structure
quantitatively consistent with the experiment, ($a-b)/(a+b)\sim0.2\%$,
while optimization without magnetism never breaks the tetragonal
symmetry. Furthermore, upon applying pressure, the hierarchy of states
changes and the single stripe state becomes the lowest in energy, as
can be seen in Fig.~\ref{fig:energies}, thus making fluctuations at
$\mathbf{q}_{\mathrm{ss}}=(\pi,0)$ the leading mode.

\vspace{0.2cm}

\textbf{Discussion}

As mentioned, there are two outstanding experimental paradoxes
regarding FeSe. The first paradox concerns the splitting of the
orthorhombic and magnetic transition observed in Fe pnictides, which
is taken to an extreme in FeSe: the structural transition occurs at
$T_s\sim 90$~K, but no magnetic order follows. Yet, exactly as in the
pnictides, DFT calculations reproduce the distorted structure when the
calculated ground state magnetic structure is used, but show no
tendency towards orbital ordering or a structural distortion if
magnetization is kept zero.

The second paradox deals with the behavior of the critical
superconducting temperature with pressure $T_c(P)$.  Typically,
pressure has a tendency to suppress magnetism, so in the context of a
magnetic pairing mechanism, pressure is beneficial to
superconductivity when magnetic order is present, but it is
destructive if it is not. For the nonmagnetic FeSe, the expectation
then is that $T_{c}$ should decrease monotonically with
pressure. Instead, $T_{c}$ first increases and then decreases with
pressure, forming a characteristic dome
shape~\cite{Medvedev2009_NatMat}.  In the following we discuss how the
{\Hb} model resolves these paradoxes.

First we analyze the {\Hb} model parameters given in Table
\ref{tab:exchangeparameters} and plotted in
Fig.~\ref{fig:phasediagram}.  The crosses in
Figs~\ref{fig:phasediagram}d, \ref{fig:phasediagram}e, and
\ref{fig:phasediagram}f, show the placement of FeSe at 9 GPa, FeSe at
0 GPa, and FeTe at 0 GPa, respectively, in the {\Hb} phase diagram.
Interestingly, for both FeSe and FeTe the calculated ground state at
ambient pressure is near a phase boundary: between the staggered dimer
phase and the single stripe phase for FeSe and between the staggered
dimer phase and double stripe phase for FeTe. Note that FeTe appears
to be very close to an Ising model because of the large $K$ and not
because of a large magnetic anisotropy.

Generally speaking, one can anticipate that, in the absence of
long-range order, spin fluctuations with wave vectors corresponding to
the lowest energy states will occur: Thus, in FeTe one expects
fluctuations with $\mathbf{q}_{\mathrm{ds}}$,
$\mathbf{q}_{\mathrm{di}}$, and $\mathbf{q}_{\mathrm{tri}}$. None of
those would support $s_{\pm }$ superconductivity since only
fluctuations with $\mathbf{q} \sim \mathbf{q}_{\mathrm{ss}}$ can pair
electrons in the standard $s_{\pm}$ superconducting state. They all
break tetragonal symmetry, but in different ways, incompatible with
each other, and cannot all support the same nematic state.  In FeSe,
by contrast, one expects fluctuations with $\mathbf{q}=(\pi ,Q)$,
where $Q=0$, $\pi /4$, $\pi /3$, and $\pi /2$ (while we cannot check
this, likely all fluctuations with $\mathbf{q}=(\pi ,Q)$, where
$ -\pi /2\lesssim Q\lesssim \pi /2$ are supported,
cf. Fig.~\ref{fig:fesl}~{\bf a}).  This is very different from Fe
pnictides, where the single stripe state is much lower in energy than
all other patterns, and therefore the $\mathbf{q}_{\mathrm{ss}}$
fluctuations dominate. Note that the above results rely upon the fact
that Fe in FeBS has a large {\it local} moment (even larger than in
DFT), \cite{Mannella2014_JPCM} and cannot be obtained by linear
response calculations based on a paramagnetic phase
\cite{Yin2014_NatPhys}.

The most important consequence of our findings is that different spin
fluctuations in FeSe (but not FeTe), while mutually incompatible with
regards to long range magnetic order, break tetragonal symmetry in the
same way (and the same is true for all $\mathbf{q}=(\pi ,Q)$,
$-\pi /2\lesssim Q\lesssim \pi /2$). In other words, one can have a
suppression of long-range magnetic ordering due to the competing
fluctuations with $\mathbf{q}=(\pi ,Q)$ for different $Q$s, but at the
same time these fluctuations all break the $x\leftrightarrow y$
symmetry and do not compete in terms of nematicity. Note that the
double-stripe fluctuations with
$\mathbf{q}_{\mathrm{ds}}=(\pi /2,\pi /2)$ break a different symmetry,
$x+y\leftrightarrow x-y$, and thus do compete nematically with the
single-stripe ones in FeTe. Therefore FeSe represents a special case
where several different types of spin fluctuations are simultaneously
excited, which prevents them from condensing at any one wave vector
and forming long range magnetic order, but does not prevent the
formation of the nematic orthorhombic order. We emphasize that this
nematic order, just as the underlying incipient magnetic one, is
accompanied by considerable orbital ordering. We find (see
Fig.~\ref{fig:fesl}~{\bf c}) that all investigated
$\mathbf{q}=(\pi ,Q)$ states induce population imbalance between the
Fe($d_{xz}$) and Fe($d_{yz}$) orbitals on each Fe site of the order of
$(n_{xz}-n_{yz})/(n_{xz}+n_{yz})\approx 8$\%. This observation
proves that the orbital ordering is $not$ sensitive to the magnetic
long range order, but only to the nematic order, and has multiple
ramifications.  The orbital ordering can be probed experimentally, and
was observed in the nematic phase at $T \lesssim T_s$ by the Knight
shift anisotropy \cite{Baek2014_NatMat}, while a divergence in
$1/TT_{1}$, as expected, was only observed at much lower temperatures,
upon approaching the long range magnetic order at $T\sim 0.$

Let us now address the intriguing pressure dependence of $T_c$.  In
general, pressure reduces magnetic interactions in FeSe.  However, the
staggered dimer state is suppressed with pressure faster than the
single stripe state (Fig.~\ref{fig:energies}), so that instead of
multiple competing types of fluctuation we obtain a situation similar
to the pnictides, where fluctuations with ${\bf q}=(\pi,0)$ decisively
dominate.  Note that in the $s_\pm$ model these are the fluctuations
that are responsible for superconductivity.  At ambient pressure, the
staggered dimer/trimer fluctuations are dominant, but cannot lead to
pairing, since the very small FeSe Fermi pockets are not connected by
${\bf q}=(\pi,Q)$, where $Q\sim \pi/2$.  As discussed in
Ref.~\onlinecite{Millis1988_PRB}, such low energy fluctuations with
``wrong'' momenta are pairbreaking since they act essentially as
impurities (note that the situation in FeSe is qualitatively different
from previous discussions in which fluctuations in different channels
compete~\cite{Fernandes2013_PRL}, but can in principle each lead to
pairing).

Under pressure the pairbreaking staggered dimer and trimer spin
fluctuations are seen to decrease in amplitude much more rapidly than
the pairing stripe spin fluctuations at ${\bf q}=(\pi,0)$.  This
removal of pairbreaking effects is responsible for the initial
increase in $T_c$. The further increase of pressure decreases the
amplitude of \textit{both} the pair-breaking $\mathbf{q} = (\pi,Q)$
and pairing ${\bf q}=(\pi,0)$ fluctuations, leading to the dome-like
behavior of $T_c$ vs. pressure.
  
Even more importantly, the nematic order, which is strongest at $P=0$,
gradually weakens with pressure, as the ${\bf q}=(\pi,Q)$
($Q\sim \pi/2$) fluctuations are suppressed. As shown in
Fig.~\ref{fig:fesl}~{\bf b}, the density of states at the Fermi level
$N(0)$ is strongly decreased in all nematic-compatible states compared
to the paramagnetic or N\'{e}el states, which is detrimental for
superconductivity (and vice-versa, as observed in
BaFe$_{2-x}$Co$_x$As$_2$ \cite{Nandi2010_PRL}). This result indicates
that long-range orbital, not magnetic, ordering leads to a sharp
reduction in the Fermi surface and thereby $N(0)$, which is consistent
with photoemission and quantum oscillation
experiments~\cite{Terashima2014_PRB,Coldea_Amalia}. Since $T_c$ is
exponentially dependent on N(0), the suppression of nematicity with
pressure is another factor ensuring the initial rise of $T_c$.

\vspace{0.2cm}

\textbf{Conclusions}

We presented a detailed analysis, based on first principles
calculations, of magnetic interactions in the FeSe/Te family. We show
that in FeSe the magnetic interactions are much more frustrated than
in either FeTe or the Fe pnictides. We argue that the simultaneous
excitation of spin fluctuations with various wave vectors of the type
$\mathbf{q}=(\pi,Q)$ prevent long-range magnetic ordering in FeSe, but
does allow for the usual spin-nematic order accompanied by a
ferro-orbital order. At zero pressure the leading fluctuations are
non-pairing (in the $s_{\pm}$ channel) $Q=\pi/2$ ones, but pairing
fluctuations at $Q=0$ become the leading fluctuations with pressure,
which explains the unusual nonmonotonic pressure dependence of $%
T_{c}$.

To be able to analyze the emerging situation on a model level, we
mapped the low-energy energetics onto a three neighbors Heisenberg +
biquadratic exchange Hamiltonian, which we have solved analytically at
$T=0$ in the mean field approximation. It appears that the biquadratic
interaction is essential to stabilize the observed double stripe phase
in FeTe; without the extra term, this phase can never be the ground
state at any choice of parameters. The same is true for the staggered
dimer phase found to be the DFT ground state in FeSe.  A nontrivial
combination of the biquadratic and third-neighbor exchanges, in
addition to the usually considered first and second neighbor
Heisenberg interactions, ensures the anomalously large splitting of
the nematic and antiferromagnetic transitions (in FeSe, it leads to a
total suppression of magnetic ordering).  We believe that this new
perspective on the unusual magnetic physics of Fe chalcogenides will
be crucial to an explanation of their remarkable properties, including
perhaps high temperature superconductivity in the monolayer FeSe
system.

\vspace{0.2cm}

\textbf{Methods}

We employed density functional theory and made use of three separate,
full potential (all electron) codes, \textsc{elk}, \textsc{wien2k},
and \textsc{fplo} to calculate the energies. The generalized gradient
approximation was used for the exchange-correlation functional. We
checked for convergence with respect to k-points and, for
\textsc{elk}, the number of empty states. We calculated the energies
of multiple collinear configurations using all three codes for
comparison purposes, while noncollinear calculations were handled
exclusively by the \textsc{elk} code. The comparison of all the
different collinear configuration energies can be found in the
Supplementary Materials.

We used the tetragonal P4/nmm space group (origin choice 2) for the
crystal structure of FeSe and FeTe in all our calculations. The Fe and
chalcogenide (Se/Te) ions occupy the 2a and 2c Wyckoff positions,
respectively. The lattice parameters for the different materials (and
for FeSe, under different pressures) are summarized in Table
\ref{tab:crystalstructure}. We note that at low temperatures FeSe is
strictly an orthorhombic structure, but this distortion is small and
omitting it leads to a small magnetoelastic error when compared with
the exchange parameter energy scales. Furthermore, we are interested
in the physics that emerges from spin fluctuations that originate in
the tetragonal phase. Therefore, for FeSe, we defined a
volume-conserving effective parameter $a^{*} = \sqrt{ab}$, where $a$
and $b$ are the orthorhombic parameters taken from experiment.

We fit to the Hamiltonian in Equation \eqref{eq:1} in the usual way.
The details of how the fit was performed are given in the
Supplementary Materials. It was not possible to achieve a fit that
accurately reproduced all energies for all possible collinear
configurations, so we defined a set of criteria for our fitting
procedure. The criteria were (1) collinear ground states of the
$J_{1} $-$ J_{2} $-$ J_{3} $-$ K$ model should be included, (2) low
energy structures that do not suffer from moment collapse under
pressure (for FeSe) should be included, (3) local moments of included
structures should be similar, and (4) we exclude configurations that
yield fits that do not reproduce the density functional theory energy
hierarchy of the lowest energy configurations. The fourth criterion is
necessary because we cannot produce an accurate fit for all
configurations, so we decide which features of the density functional
theory set of energies is important from the point of view of
fluctuations and frustration, which are the lowest energy ones.  Given
these criteria, we perform the fitting procedure using the energies
summarized in Fig.~\ref{fig:energies}.

\vspace{0.2cm}

\textbf{Acknowledgments}

We thank M. Tomi{\'c} for running some test calculations at the
initial stages of this work and D. Guterding, S. Backes, A. Coldea,
R. Fernandes, N. Perkins, S. Kivelson and Wei Ku for valuable
discussions.  I.I.M. is supported by ONR through the NRL basic
research program. J.K.G. acknowledges the support of the NRC program
at NRL. H.O.J. and R.V. are supported by DFG-SPP1458. P.J.H. was
partially supported by US DOE DE-FG02-05ER46236.

\vspace{0.2cm}

\textbf{Author contributions}

I.I.M. and J.K.G. conceived the research; J.K.G., I.I.M. and
H.O.J. carried out numerical calculations; all authors participated in
the discussion and contributed to writing the paper; I.I.M. and
R.V. supervised the whole project.

\appendix

\section*{Supplementary Material}
\label{sec:suppl-mater}

\textbf{Supplementary Methods}

We calculated the energy of a variety of different collinear
structures using the three different codes, \textsc{elk} \cite{elk},
\textsc{wien2k} \cite{wien2k}, and \textsc{fplo}
\cite{Koepernik1999_PRB}. The generalized gradient approximation was
used for the exchange-correlation functional
\cite{Perdew1996_PRL}. The structures in
Fig.~\ref{fig:suppmagstructures} summarize all of the different
configurations that we considered. In Fig.~\ref{fig:magenergies} are
the energies we calculated using these codes. Note that we did not
calculate the energy of every configuration using all three codes, but
there are several points of comparison. For all configurations for
which we can make a comparison, there is excellent agreement across
codes. The most important result of this comparison is that there is
no ambiguity as to the energy hierarchy of the low-lying energy
states, it is the same for all three codes. We also note that the
energy range for the different configurations is quite large for both
FeSe and FeTe, on the scale of $100-300$ meV for FeSe and $50-100$ meV
for FeTe.

We fitted to the Heisenberg model with the ordinary least squares
(OLS) method using the Heisenberg model coefficients reported in Table
\ref{tab:heisenbergcoefficients} for our collinear fits and the
expression
$\Delta E(\theta) = E(\theta) - E(0) = 2 K \sin^{2}(\theta)$, see
Ref.~\cite{Glasbrenner2014b_PRB} for the definition of $\theta$, for
our noncollinear fits. The noncollinear energies and the corresponding
fits are shown in Fig.~\ref{fig:noncollinear_energy}.

As reported in the main article, the collinear fits are very good for
the included configurations, but there are deviations if we apply the
model to configurations excluded due to the criteria we outlined in
the main Methods section. This is a consequence of the itinerant
nature of the magnetism, which in general cannot be mapped onto a
pairwise interaction model. We also note that the lower symmetry
magnetic structures, such as those with generic names such as
``dduuduuu,'' suffered moment collapse in FeSe under pressure. The
noncollinear fits, on the other hand, are excellent.

It is worth noting that for itinerant magnets the exchange model could
be potentially improved using an approach similar to Moriya
\cite{Moriya1985_Book} and allowing the moment amplitudes to vary and
by also including Stoner-like onsite terms. We tried including terms
like this to see how it affected the quality of our fits. We found
that including these terms does not change the fitting results in any
qualitative way when using the configurations in Table
\ref{tab:heisenbergcoefficients}. Furthermore, it did not allow us to
extend the fit to also reproduce the high-energy configurations from
Fig.~\ref{fig:suppmagstructures}.  It is possible, however, that these
modifications would be important for fluctuations above the N\'{e}el
temperature.

\vspace{0.2cm}

\textbf{Phase boundaries of $J_1$-$J_2$-$J_3$-$K$ model}

Here we give additional details of the analytic solution of the
$J_1$-$J_2$-$J_3$-$K$ model. First, let us ignore the $K$ term and
refresh what is known about the $J_{1}-J_{2}-J_{3}$ Heisenberg
models. The Ising model has four phases, the checkerboard (cb) phase
in Fig.~\ref{fig:suppmagstructures}a, the double stripe (ds) phase in
Fig.~\ref{fig:suppmagstructures}d, the single stripe (ss) phase in
Fig.~\ref{fig:suppmagstructures}e, and the staggered dimers (di) phase
in Fig.~\ref{fig:suppmagstructures}f. The Heisenberg model also has
four phases, but neither the ds or di phase are ground states in the
phase diagram. Instead, the four phases are the aforementioned cb and
ss phases, and in addition two spiral phases with spins rotating away
from the origin as $\alpha = n_{x}q_{x} + n_{y}q_{y}$, the first with
wavevector $\mathbf{q}_{1} = (\pi,Q)$ and the second with
$\mathbf{q}_{2} = (Q,Q)$ (see, e.g.,
Ref.~\cite{Pimpinelli1990_PRB}). Note that at
$\mathbf{q}_{1}(Q \to 0) = (\pi,0)$, which is the ss phase, and at
$\mathbf{q}_{1}(Q \to \pi) = \mathbf{q}_{2}(Q \to \pi) = (\pi,\pi)$,
which is the cb phase. In both phases $Q$ depends on the exchange
parameters:
$Q = \cos^{-1} \left[\left(2J_{2}-J_{1}\right)/4J_{3}\right]$ for
$\mathbf{q}_{1}$ and
$Q = \pi - \cos^{-1}\left[J_{1}/\left(2J_{2}+4J_{3}\right)\right]$ for
$\mathbf{q}_{2}$. Finally, the analytic expressions for the phase
boundaries are summarized in Table \ref{tab:phaseboundaries}.

Adding in the biquadratic term
$-K \left(\hat{\mathbf{m}}_{i} \cdot \hat{\mathbf{m}}_{j}\right)^{2}$
restores the ds and di configurations to the phase diagram. The
allowed wavevectors in the spin spiral phases also become dependent on
$K$:
$Q = \cos^{-1} \left[\left(2J_{2}-J_{1}\right)/\left(4J_{3} -
    2K\right) \right]$
for $\mathbf{q}_{1}$ and
$Q = \pi - \cos^{-1}\left[J_{1}/\left(2J_{2}+4J_{3}-2K\right)\right]$
for $\mathbf{q}_{2}$. The analytic expressions for the phase
boundaries also change and many become $K$-dependent as summarized in
the last column of Table \ref{tab:phaseboundaries}. As $K$ grows so do
the areas of stability of the ds and di phases. Once $K > J_{1}/2$,
the phase diagram becomes indistinguishable from the Ising model.

\vspace{0.2cm}
 
\textbf{FeSe$_{0.5}$Te$_{0.5}$}

A notable omission to our results is the case of
FeSe$_{0.50}$Te$_{0.50}$. Unlike the other materials, the structure of
FeSe$_{0.50}$Te$_{0.50}$ is not well-defined. A common approach is to
use lattice parameters from experiment and then choose the
chalcogenide to be either pure Se or pure Te, assuming that the change
in the lattice parameters drives the relevant physics, such as
inducing superconductivity. A check of this reveals that this is not
entirely the case; there is a significant energy splitting of the
checkerboard, double stripe, and zig-zag configurations when Se/Te are
swapped, and the energy splits are not in the same direction. Using Te
lowers the checkerboard energy, while it increases the double stripe
energy, for example. Furthermore, careful experimental analysis
reveals that FeSe$_{0.50}$Te$_{0.50}$ is a disordered structure with
different heights for Se and Te \cite{Louca2010_PRB}. Taking this into
account requires an expensive and non-trivial averaging
procedure. While a description of FeSe$_{0.50}$Te$_{0.50}$ would be
useful, we put the question aside for now due to the complexity of the
structure.

\bibliographystyle{naturemag}
\bibliography{fese,fese_supplement}

\begin{thebibliography}{10}
\expandafter\ifx\csname url\endcsname\relax
  \def\url#1{\texttt{#1}}\fi
\expandafter\ifx\csname urlprefix\endcsname\relax\def\urlprefix{URL }\fi
\providecommand{\bibinfo}[2]{#2}
\providecommand{\eprint}[2][]{\url{#2}}

\bibitem{Chubukov2012_ARCMP}
\bibinfo{author}{Chubukov, A.}
\newblock \bibinfo{title}{{Pairing Mechanism in Fe-Based Superconductors}}.
\newblock \emph{\bibinfo{journal}{Annu. Rev. Condens. Matter Phys.}}
  \textbf{\bibinfo{volume}{3}}, \bibinfo{pages}{57} (\bibinfo{year}{2012}).

\bibitem{Hirschfeld2011_RPP}
\bibinfo{author}{Hirschfeld, P.~J.}, \bibinfo{author}{Korshunov, M.~M.} \&
  \bibinfo{author}{Mazin, I.~I.}
\newblock \bibinfo{title}{{Gap symmetry and structure of Fe-based
  superconductors}}.
\newblock \emph{\bibinfo{journal}{Rep. Prog. Phys.}}
  \textbf{\bibinfo{volume}{74}}, \bibinfo{pages}{124508}
  (\bibinfo{year}{2011}).

\bibitem{Wang2009_EPL}
\bibinfo{author}{Wang, F.}, \bibinfo{author}{Zhai, H.} \& \bibinfo{author}{Lee,
  D.-H.}
\newblock \bibinfo{title}{{Antiferromagnetic correlation and the pairing
  mechanism of the cuprates and iron pnictides: A view from the functional
  renormalization group studies}}.
\newblock \emph{\bibinfo{journal}{Europhys. Lett.}}
  \textbf{\bibinfo{volume}{85}}, \bibinfo{pages}{37005} (\bibinfo{year}{2009}).

\bibitem{Chubukov2008_PRB}
\bibinfo{author}{Chubukov, A.~V.}, \bibinfo{author}{Efremov, D.~V.} \&
  \bibinfo{author}{Eremin, I.}
\newblock \bibinfo{title}{{Magnetism, superconductivity, and pairing symmetry
  in iron-based superconductors}}.
\newblock \emph{\bibinfo{journal}{Phys. Rev. B}} \textbf{\bibinfo{volume}{78}},
  \bibinfo{pages}{134512} (\bibinfo{year}{2008}).

\bibitem{Stanev2008_PRB}
\bibinfo{author}{Stanev, V.}, \bibinfo{author}{Kang, J.} \&
  \bibinfo{author}{Tesanovic, Z.}
\newblock \bibinfo{title}{{Spin fluctuation dynamics and multiband
  superconductivity in iron pnictides}}.
\newblock \emph{\bibinfo{journal}{Phys. Rev. B}} \textbf{\bibinfo{volume}{78}},
  \bibinfo{pages}{184509} (\bibinfo{year}{2008}).

\bibitem{Kuroki2008_PRL}
\bibinfo{author}{Kuroki, K.} \emph{et~al.}
\newblock \bibinfo{title}{{Unconventional Pairing Originating from the
  Disconnected Fermi Surfaces of Superconducting
  \mbox{LaFeAsO}$_{1-x}$\mbox{F}$_{x}$}}.
\newblock \emph{\bibinfo{journal}{Phys. Rev. Lett.}}
  \textbf{\bibinfo{volume}{101}}, \bibinfo{pages}{087004}
  (\bibinfo{year}{2008}).

\bibitem{Graser2009_NJP}
\bibinfo{author}{Graser, S.}, \bibinfo{author}{Maier, T.~A.},
  \bibinfo{author}{Hirschfeld, P.~J.} \& \bibinfo{author}{Scalapino, D.~J.}
\newblock \bibinfo{title}{{Near-degeneracy of several pairing channels in
  multiorbital models for the Fe pnictides}}.
\newblock \emph{\bibinfo{journal}{New J. Phys.}} \textbf{\bibinfo{volume}{11}},
  \bibinfo{pages}{025016} (\bibinfo{year}{2009}).

\bibitem{Sknepnek2009_PRB}
\bibinfo{author}{Sknepnek, R.}, \bibinfo{author}{Samolyuk, G.},
  \bibinfo{author}{Lee, Y.-B.} \& \bibinfo{author}{Schmalian, J.}
\newblock \bibinfo{title}{{Anisotropy of the pairing gap of FeAs-based
  superconductors induced by spin fluctuations}}.
\newblock \emph{\bibinfo{journal}{Phys. Rev. B}} \textbf{\bibinfo{volume}{79}},
  \bibinfo{pages}{054511} (\bibinfo{year}{2009}).

\bibitem{Yao2009_NJP}
\bibinfo{author}{Yao, Z.-J.}, \bibinfo{author}{Li, J.-X.} \&
  \bibinfo{author}{Wang, Z.~D.}
\newblock \bibinfo{title}{{Spin fluctuations, interband coupling and
  unconventional pairing in iron-based superconductors}}.
\newblock \emph{\bibinfo{journal}{New J. Phys.}} \textbf{\bibinfo{volume}{11}},
  \bibinfo{pages}{025009} (\bibinfo{year}{2009}).

\bibitem{Guterding2015_PRB}
\bibinfo{author}{Guterding, D.}, \bibinfo{author}{Jeschke, H.~O.},
  \bibinfo{author}{Hirschfeld, P.~J.} \& \bibinfo{author}{Valent\'\i, R.}
\newblock \bibinfo{title}{{Unified picture of the doping dependence of
  superconducting transition temperatures in alkali metal/ammonia intercalated
  FeSe}}.
\newblock \emph{\bibinfo{journal}{Phys. Rev. B}} \textbf{\bibinfo{volume}{91}},
  \bibinfo{pages}{041112} (\bibinfo{year}{2015}).

\bibitem{Seo2008_PRL}
\bibinfo{author}{Seo, K.}, \bibinfo{author}{Bernevig, B.~A.} \&
  \bibinfo{author}{Hu, J.}
\newblock \bibinfo{title}{{Pairing Symmetry in a Two-Orbital Exchange Coupling
  Model of Oxypnictides}}.
\newblock \emph{\bibinfo{journal}{Phys. Rev. Lett.}}
  \textbf{\bibinfo{volume}{101}}, \bibinfo{pages}{206404}
  (\bibinfo{year}{2008}).

\bibitem{Lv2010_PRB}
\bibinfo{author}{Lv, W.}, \bibinfo{author}{Kr\"uger, F.} \&
  \bibinfo{author}{Phillips, P.}
\newblock \bibinfo{title}{{Orbital ordering and unfrustrated $(\pi,0)$
  magnetism from degenerate double exchange in the iron pnictides}}.
\newblock \emph{\bibinfo{journal}{Phys. Rev. B}} \textbf{\bibinfo{volume}{82}},
  \bibinfo{pages}{045125} (\bibinfo{year}{2010}).

\bibitem{Dai2012_NatPhys}
\bibinfo{author}{Dai, P.}, \bibinfo{author}{Hu, J.} \&
  \bibinfo{author}{Dagotto, E.}
\newblock \bibinfo{title}{{Magnetism and its microscopic origin in iron-based
  high-temperature superconductors}}.
\newblock \emph{\bibinfo{journal}{Nat. Phys.}} \textbf{\bibinfo{volume}{8}},
  \bibinfo{pages}{709} (\bibinfo{year}{2012}).

\bibitem{Moon2010_PRB}
\bibinfo{author}{Moon, S.~J.} \emph{et~al.}
\newblock \bibinfo{title}{{Dual character of magnetism in
  \mbox{EuFe}$_{2}$\mbox{As}$_{2}$: Optical spectroscopic and
  density-functional calculation study}}.
\newblock \emph{\bibinfo{journal}{Phys. Rev. B}} \textbf{\bibinfo{volume}{81}},
  \bibinfo{pages}{205114} (\bibinfo{year}{2010}).

\bibitem{Lee2010_PRB}
\bibinfo{author}{Lee, H.}, \bibinfo{author}{Zhang, Y.-Z.},
  \bibinfo{author}{Jeschke, H.~O.} \& \bibinfo{author}{Valent\'\i, R.}
\newblock \bibinfo{title}{{Possible origin of the reduced ordered magnetic
  moment in iron pnictides: A dynamical mean-field theory study}}.
\newblock \emph{\bibinfo{journal}{Phys. Rev. B}} \textbf{\bibinfo{volume}{81}},
  \bibinfo{pages}{220506} (\bibinfo{year}{2010}).

\bibitem{Yin2011_NatMater}
\bibinfo{author}{Yin, Z.~P.}, \bibinfo{author}{Haule, K.} \&
  \bibinfo{author}{Kotliar, G.}
\newblock \bibinfo{title}{{Kinetic frustration and the nature of the magnetic
  and paramagnetic states in iron pnictides and iron chalcogenides}}.
\newblock \emph{\bibinfo{journal}{Nat. Mater.}} \textbf{\bibinfo{volume}{10}},
  \bibinfo{pages}{932} (\bibinfo{year}{2011}).

\bibitem{Chandra1990_PRL}
\bibinfo{author}{Chandra, P.}, \bibinfo{author}{Coleman, P.} \&
  \bibinfo{author}{Larkin, A.~I.}
\newblock \bibinfo{title}{{Ising transition in frustrated Heisenberg models}}.
\newblock \emph{\bibinfo{journal}{Phys. Rev. Lett.}}
  \textbf{\bibinfo{volume}{64}}, \bibinfo{pages}{88} (\bibinfo{year}{1990}).

\bibitem{Xu2008_PRB}
\bibinfo{author}{Xu, C.}, \bibinfo{author}{M\"uller, M.} \&
  \bibinfo{author}{Sachdev, S.}
\newblock \bibinfo{title}{{Ising and spin orders in the iron-based
  superconductors}}.
\newblock \emph{\bibinfo{journal}{Phys. Rev. B}} \textbf{\bibinfo{volume}{78}},
  \bibinfo{pages}{020501} (\bibinfo{year}{2008}).

\bibitem{Fang2008_PRB}
\bibinfo{author}{Fang, C.}, \bibinfo{author}{Yao, H.}, \bibinfo{author}{Tsai,
  W.-F.}, \bibinfo{author}{Hu, J.} \& \bibinfo{author}{Kivelson, S.~A.}
\newblock \bibinfo{title}{{Theory of electron nematic order in LaFeAsO}}.
\newblock \emph{\bibinfo{journal}{Phys. Rev. B}} \textbf{\bibinfo{volume}{77}},
  \bibinfo{pages}{224509} (\bibinfo{year}{2008}).

\bibitem{Ma2008_PRB}
\bibinfo{author}{Ma, F.}, \bibinfo{author}{Lu, Z.-Y.} \&
  \bibinfo{author}{Xiang, T.}
\newblock \bibinfo{title}{{Arsenic-bridged antiferromagnetic superexchange
  interactions in LaFeAsO}}.
\newblock \emph{\bibinfo{journal}{Phys. Rev. B}} \textbf{\bibinfo{volume}{78}},
  \bibinfo{pages}{224517} (\bibinfo{year}{2008}).

\bibitem{Si2008_PRL}
\bibinfo{author}{Si, Q.} \& \bibinfo{author}{Abrahams, E.}
\newblock \bibinfo{title}{{Strong Correlations and Magnetic Frustration in the
  High ${T}_{c}$ Iron Pnictides}}.
\newblock \emph{\bibinfo{journal}{Phys. Rev. Lett.}}
  \textbf{\bibinfo{volume}{101}}, \bibinfo{pages}{076401}
  (\bibinfo{year}{2008}).

\bibitem{Zhao2009_NatPhys}
\bibinfo{author}{Zhao, J.} \emph{et~al.}
\newblock \bibinfo{title}{{Spin waves and magnetic exchange interactions in
  CaFe$_2$As$_2$}}.
\newblock \emph{\bibinfo{journal}{Nat. Phys.}} \textbf{\bibinfo{volume}{5}},
  \bibinfo{pages}{555} (\bibinfo{year}{2009}).

\bibitem{Diallo2009_PRL}
\bibinfo{author}{Diallo, S.~O.} \emph{et~al.}
\newblock \bibinfo{title}{{Itinerant Magnetic Excitations in Antiferromagnetic
  \mbox{CaFe}$_{2}$\mbox{As}$_{2}$}}.
\newblock \emph{\bibinfo{journal}{Phys. Rev. Lett.}}
  \textbf{\bibinfo{volume}{102}}, \bibinfo{pages}{187206}
  (\bibinfo{year}{2009}).

\bibitem{Yaresko2009_PRB}
\bibinfo{author}{Yaresko, A.~N.}, \bibinfo{author}{Liu, G.-Q.},
  \bibinfo{author}{Antonov, V.~N.} \& \bibinfo{author}{Andersen, O.~K.}
\newblock \bibinfo{title}{{Interplay between magnetic properties and Fermi
  surface nesting in iron pnictides}}.
\newblock \emph{\bibinfo{journal}{Phys. Rev. B}} \textbf{\bibinfo{volume}{79}},
  \bibinfo{pages}{144421} (\bibinfo{year}{2009}).

\bibitem{Li2009_PRB}
\bibinfo{author}{Li, S.} \emph{et~al.}
\newblock \bibinfo{title}{{First-order magnetic and structural phase
  transitions in Fe$_{1+y}$Se$_{x}$Te$_{1+x}$}}.
\newblock \emph{\bibinfo{journal}{Phys. Rev. B}} \textbf{\bibinfo{volume}{79}},
  \bibinfo{pages}{054503} (\bibinfo{year}{2009}).

\bibitem{Bao2009_PRL}
\bibinfo{author}{Bao, W.} \emph{et~al.}
\newblock \bibinfo{title}{{Tunable $(\delta\pi, \delta\pi)$-Type
  Antiferromagnetic Order in $\alpha$-Fe(Te,Se) Superconductors}}.
\newblock \emph{\bibinfo{journal}{Phys. Rev. Lett.}}
  \textbf{\bibinfo{volume}{102}}, \bibinfo{pages}{247001}
  (\bibinfo{year}{2009}).

\bibitem{Ma2009_PRL}
\bibinfo{author}{Ma, F.}, \bibinfo{author}{Ji, W.}, \bibinfo{author}{Hu, J.},
  \bibinfo{author}{Lu, Z.-Y.} \& \bibinfo{author}{Xiang, T.}
\newblock \bibinfo{title}{{First-Principles Calculations of the Electronic
  Structure of Tetragonal \(\alpha\)-FeTe and \(\alpha\)-FeSe Crystals:
  Evidence for a Bicollinear Antiferromagnetic Order}}.
\newblock \emph{\bibinfo{journal}{Phys. Rev. Lett.}}
  \textbf{\bibinfo{volume}{102}}, \bibinfo{pages}{177003}
  (\bibinfo{year}{2009}).

\bibitem{Ferrer1993_PRB}
\bibinfo{author}{Ferrer, J.}
\newblock \bibinfo{title}{{Spin-liquid phase for the frustrated quantum
  Heisenberg antiferromagnet on a square lattice}}.
\newblock \emph{\bibinfo{journal}{Phys. Rev. B}} \textbf{\bibinfo{volume}{47}},
  \bibinfo{pages}{8769} (\bibinfo{year}{1993}).

\bibitem{Sindzingre2010_JPCS}
\bibinfo{author}{Sindzingre, P.}, \bibinfo{author}{Shannon, N.} \&
  \bibinfo{author}{Momoi, T.}
\newblock \bibinfo{title}{{Phase diagram of the spin-1/2 $J_1-J_2-J_3$
  Heisenberg model on the square lattice}}.
\newblock \emph{\bibinfo{journal}{J. Phys. Conf. Ser.}}
  \textbf{\bibinfo{volume}{200}}, \bibinfo{pages}{022058}
  (\bibinfo{year}{2010}).

\bibitem{Wysocki2011_NatPhys}
\bibinfo{author}{Wysocki, A.~L.}, \bibinfo{author}{Belashchenko, K.~D.} \&
  \bibinfo{author}{Antropov, V.~P.}
\newblock \bibinfo{title}{{Consistent model of magnetism in ferropnictides}}.
\newblock \emph{\bibinfo{journal}{Nat. Phys.}} \textbf{\bibinfo{volume}{7}},
  \bibinfo{pages}{485} (\bibinfo{year}{2011}).

\bibitem{Wysocki2013_JPCS}
\bibinfo{author}{Wysocki, A.~L.}, \bibinfo{author}{Belashchenko, K.~D.},
  \bibinfo{author}{Ke, L.}, \bibinfo{author}{van Schilfgaarde, M.} \&
  \bibinfo{author}{Antropov, V.~P.}
\newblock \bibinfo{title}{{Biquadratic magnetic interaction in parent
  ferropnictides}}.
\newblock \emph{\bibinfo{journal}{J. Phys. Conf. Ser.}}
  \textbf{\bibinfo{volume}{449}}, \bibinfo{pages}{012024}
  (\bibinfo{year}{2013}).

\bibitem{Hu2012_PRB}
\bibinfo{author}{Hu, J.}, \bibinfo{author}{Xu, B.}, \bibinfo{author}{Liu, W.},
  \bibinfo{author}{Hao, N.-N.} \& \bibinfo{author}{Wang, Y.}
\newblock \bibinfo{title}{{Unified minimum effective model of magnetic
  properties of iron-based superconductors}}.
\newblock \emph{\bibinfo{journal}{Phys. Rev. B}} \textbf{\bibinfo{volume}{85}},
  \bibinfo{pages}{144403} (\bibinfo{year}{2012}).

\bibitem{Fernandes2012_PRB}
\bibinfo{author}{Fernandes, R.~M.}, \bibinfo{author}{Chubukov, A.~V.},
  \bibinfo{author}{Knolle, J.}, \bibinfo{author}{Eremin, I.} \&
  \bibinfo{author}{Schmalian, J.}
\newblock \bibinfo{title}{{Preemptive nematic order, pseudogap, and orbital
  order in the iron pnictides}}.
\newblock \emph{\bibinfo{journal}{Phys. Rev. B}} \textbf{\bibinfo{volume}{85}},
  \bibinfo{pages}{024534} (\bibinfo{year}{2012}).

\bibitem{Mazin2009_PhysicaC}
\bibinfo{author}{Mazin, I.~I.} \& \bibinfo{author}{Schmalian, J.}
\newblock \bibinfo{title}{{Pairing symmetry and pairing state in
  ferropnictides: Theoretical overview}}.
\newblock \emph{\bibinfo{journal}{Physica C}} \textbf{\bibinfo{volume}{469}},
  \bibinfo{pages}{614} (\bibinfo{year}{2009}).

\bibitem{Baek2014_NatMat}
\bibinfo{author}{Baek, S.-H.} \emph{et~al.}
\newblock \bibinfo{title}{{Orbital-driven nematicity in FeSe}}.
\newblock \emph{\bibinfo{journal}{Nat. Mater.}}  (\bibinfo{year}{2014}).

\bibitem{Bohmer2015_PRL}
\bibinfo{author}{B{\"o}hmer, A.~E.} \emph{et~al.}
\newblock \bibinfo{title}{{Origin of the Tetragonal-to-Orthorhombic Phase
  Transition in FeSe: A Combined Thermodynamic and NMR Study of Nematicity}}.
\newblock \emph{\bibinfo{journal}{Phys. Rev. Lett.}}
  \textbf{\bibinfo{volume}{114}}, \bibinfo{pages}{027001}
  (\bibinfo{year}{2015}).

\bibitem{Medvedev2009_NatMat}
\bibinfo{author}{Medvedev, S.} \emph{et~al.}
\newblock \bibinfo{title}{{Electronic and magnetic phase diagram of
  $\beta$-Fe$_{1.01}$Se with superconductivity at 36.7 K under pressure}}.
\newblock \emph{\bibinfo{journal}{Nat. Mater.}} \textbf{\bibinfo{volume}{8}},
  \bibinfo{pages}{630} (\bibinfo{year}{2009}).

\bibitem{Landau1985_PRB}
\bibinfo{author}{Landau, D.~P.} \& \bibinfo{author}{Binder, K.}
\newblock \bibinfo{title}{{Phase diagrams and critical behavior of Ising square
  lattices with nearest-, next-nearest-, and third-nearest-neighbor
  couplings}}.
\newblock \emph{\bibinfo{journal}{Phys. Rev. B}} \textbf{\bibinfo{volume}{31}},
  \bibinfo{pages}{5946} (\bibinfo{year}{1985}).

\bibitem{Cao2014_arXiv}
\bibinfo{author}{Cao, H.-Y.}, \bibinfo{author}{Chen, S.},
  \bibinfo{author}{Xiang, H.} \& \bibinfo{author}{Gong, X.-G.}
\newblock \bibinfo{title}{{Antiferromagnetic ground state with pair-checkboard
  order in FeSe}}.
\newblock \eprint{Preprint at <http://arxiv.org/abs/1407.7145>}.

\bibitem{Glasbrenner2014b_PRB}
\bibinfo{author}{Glasbrenner, J.~K.}, \bibinfo{author}{Velev, J.~P.} \&
  \bibinfo{author}{Mazin, I.~I.}
\newblock \bibinfo{title}{{First-principles study of the minimal model of
  magnetic interactions in Fe-based superconductors}}.
\newblock \emph{\bibinfo{journal}{Phys. Rev. B}} \textbf{\bibinfo{volume}{89}},
  \bibinfo{pages}{064509} (\bibinfo{year}{2014}).

\bibitem{Mannella2014_JPCM}
\bibinfo{author}{Mannella, N.}
\newblock \bibinfo{title}{{The magnetic moment enigma in Fe-based high
  temperature superconductors}}.
\newblock \emph{\bibinfo{journal}{J. Phys. Condens. Mat.}}
  \textbf{\bibinfo{volume}{26}}, \bibinfo{pages}{473202}
  (\bibinfo{year}{2014}).

\bibitem{Yin2014_NatPhys}
\bibinfo{author}{Yin, Z.~P.}, \bibinfo{author}{Haule, K.} \&
  \bibinfo{author}{Kotliar, G.}
\newblock \bibinfo{title}{{Spin dynamics and orbital-antiphase pairing symmetry
  in iron-based superconductors}}.
\newblock \emph{\bibinfo{journal}{Nat. Phys.}} \textbf{\bibinfo{volume}{10}},
  \bibinfo{pages}{845} (\bibinfo{year}{2014}).

\bibitem{Millis1988_PRB}
\bibinfo{author}{Millis, A.~J.}, \bibinfo{author}{Sachdev, S.} \&
  \bibinfo{author}{Varma, C.~M.}
\newblock \bibinfo{title}{{Inelastic scattering and pair breaking in
  anisotropic and isotropic superconductors}}.
\newblock \emph{\bibinfo{journal}{Phys. Rev. B}} \textbf{\bibinfo{volume}{37}},
  \bibinfo{pages}{4975} (\bibinfo{year}{1988}).

\bibitem{Fernandes2013_PRL}
\bibinfo{author}{Fernandes, R.~M.} \& \bibinfo{author}{Millis, A.~J.}
\newblock \bibinfo{title}{{Nematicity as a Probe of Superconducting Pairing in
  Iron-Based Superconductors}}.
\newblock \emph{\bibinfo{journal}{Phys. Rev. Lett.}}
  \textbf{\bibinfo{volume}{111}}, \bibinfo{pages}{127001}
  (\bibinfo{year}{2013}).

\bibitem{Nandi2010_PRL}
\bibinfo{author}{Nandi, S.} \emph{et~al.}
\newblock \bibinfo{title}{{Anomalous Suppression of the Orthorhombic Lattice
  Distortion in Superconducting
  \mbox{Ba}(\mbox{Fe}$_{1-x}$\mbox{Co}$_{x}$)$_{2}$\mbox{As}$_{2}$ Single
  Crystals}}.
\newblock \emph{\bibinfo{journal}{Phys. Rev. Lett.}}
  \textbf{\bibinfo{volume}{104}}, \bibinfo{pages}{057006}
  (\bibinfo{year}{2010}).

\bibitem{Terashima2014_PRB}
\bibinfo{author}{Terashima, T.} \emph{et~al.}
\newblock \bibinfo{title}{{Anomalous Fermi surface in FeSe seen by Shubnikov-de
  Haas oscillation measurements}}.
\newblock \emph{\bibinfo{journal}{Phys. Rev. B}} \textbf{\bibinfo{volume}{90}},
  \bibinfo{pages}{144517} (\bibinfo{year}{2014}).

\bibitem{Coldea_Amalia}
\bibinfo{note}{Coldea, A. Private communication.}

\bibitem{elk}
\bibinfo{note}{{ELK FP-LAPW Code [http://elk.sourceforge.net/]}}.

\bibitem{wien2k}
\bibinfo{author}{Blaha, P.}, \bibinfo{author}{Schwarz, K.},
  \bibinfo{author}{Madsen, G. K.~H.}, \bibinfo{author}{Kvasnicka, D.} \&
  \bibinfo{author}{Luitz, J.}
\newblock \emph{\bibinfo{title}{{WIEN2k, An Augmented Plane Wave + Local
  Orbitals Program for Calculating Crystal Properties}}}
  (\bibinfo{publisher}{Techn. Universit\"at Wien, Austria},
  \bibinfo{year}{2001}).

\bibitem{Koepernik1999_PRB}
\bibinfo{author}{Koepernik, K.} \& \bibinfo{author}{Eschrig, H.}
\newblock \bibinfo{title}{{Full-potential nonorthogonal local-orbital
  minimum-basis band-structure scheme}}.
\newblock \emph{\bibinfo{journal}{Phys. Rev. B}} \textbf{\bibinfo{volume}{59}},
  \bibinfo{pages}{1743} (\bibinfo{year}{1999}).

\bibitem{Perdew1996_PRL}
\bibinfo{author}{Perdew, J.~P.}, \bibinfo{author}{Burke, K.} \&
  \bibinfo{author}{Ernzerhof, M.}
\newblock \bibinfo{title}{{Generalized Gradient Approximation Made Simple}}.
\newblock \emph{\bibinfo{journal}{Phys. Rev. Lett.}}
  \textbf{\bibinfo{volume}{77}}, \bibinfo{pages}{3865} (\bibinfo{year}{1996}).

\bibitem{Moriya1985_Book}
\bibinfo{author}{Moriya, T.}
\newblock \emph{\bibinfo{title}{{Spin Fluctuations in Itinerant Electron
  Magnetism}}} (\bibinfo{publisher}{Springer}, \bibinfo{address}{Berlin},
  \bibinfo{year}{1985}).

\bibitem{Pimpinelli1990_PRB}
\bibinfo{author}{Pimpinelli, A.} \& \bibinfo{author}{Rastelli, E.}
\newblock \bibinfo{title}{{Absence of long-range order in three-dimensional
  spherical models}}.
\newblock \emph{\bibinfo{journal}{Phys. Rev. B}} \textbf{\bibinfo{volume}{42}},
  \bibinfo{pages}{984} (\bibinfo{year}{1990}).

\bibitem{Louca2010_PRB}
\bibinfo{author}{Louca, D.} \emph{et~al.}
\newblock \bibinfo{title}{{Local atomic structure of superconducting
  FeSe$_{1-x}$Te$_{x}$}}.
\newblock \emph{\bibinfo{journal}{Phys. Rev. B}} \textbf{\bibinfo{volume}{81}},
  \bibinfo{pages}{134524} (\bibinfo{year}{2010}).

\bibitem{Margadonna2009_PRB}
\bibinfo{author}{Margadonna, S.} \emph{et~al.}
\newblock \bibinfo{title}{{Pressure evolution of the low-temperature crystal
  structure and bonding of the superconductor FeSe $({T}_{c}=37\text{
  }\text{K})$}}.
\newblock \emph{\bibinfo{journal}{Phys. Rev. B}} \textbf{\bibinfo{volume}{80}},
  \bibinfo{pages}{064506} (\bibinfo{year}{2009}).

\end{thebibliography}

\newpage 

\begin{figure*}[t]
\centering
\includegraphics[width=0.98\textwidth]{allphases}
\caption{\textbf{Classical mean-field phase diagrams.} \textbf{a}, The
  $J_1$-$J_2$-$J_3$ Ising model. \textbf{b}, The $J_1$-$J_2$-$J_3$
  Heisenberg model. \textbf{c}, The {\Hb} model
  with $K=0.1$. \textbf{d}, The {\Hb} model with
  $K=0.20$ where the cross corresponds to FeSe at 9 GPa of
  pressure. \textbf{e}, The {\Hb} model with
  $K=0.25$ where the cross corresponds to FeSe at 0 GPa of
  pressure. \textbf{f}, The {\Hb} model with
  $K=0.39$ where the cross corresponds to FeTe at 0 GPa of
  pressure. The length of the cross' bars in panels \textbf{d} through
  \textbf{f} indicate the uncertainty of the fit to the {\Hb} model.}
\label{fig:phasediagram}
\end{figure*}

\begin{figure*}[t]
\centering
\includegraphics[width=0.48\textwidth]{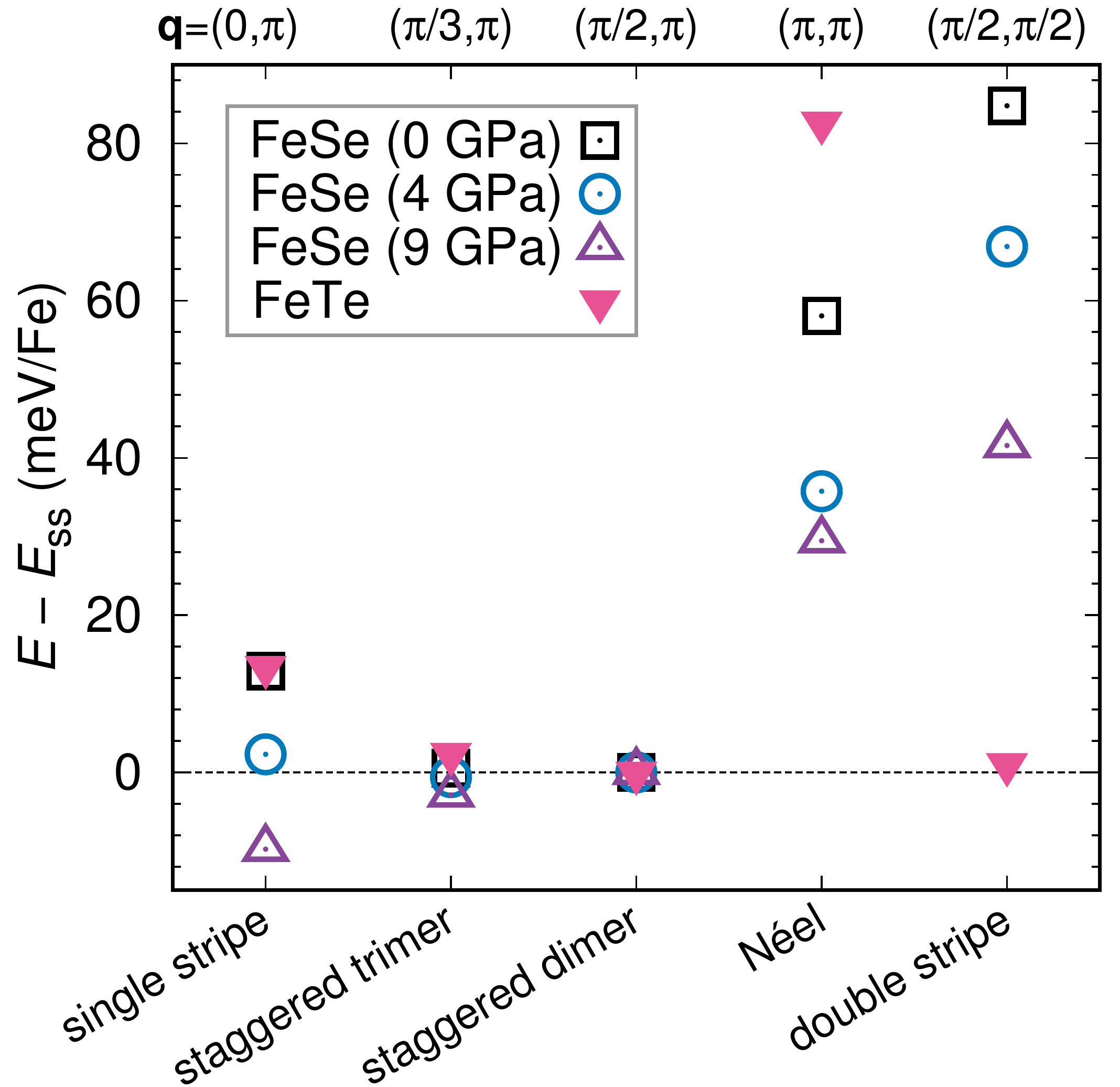}
\caption{\textbf{Energies of collinear magnetic configurations.} The
  total calculated DFT energies of the collinear magnetic
  configurations used in the fits of the {\Hb} model.}
\label{fig:energies}
\end{figure*}

\begin{figure*}[t]
  \includegraphics[width=0.75\textwidth]{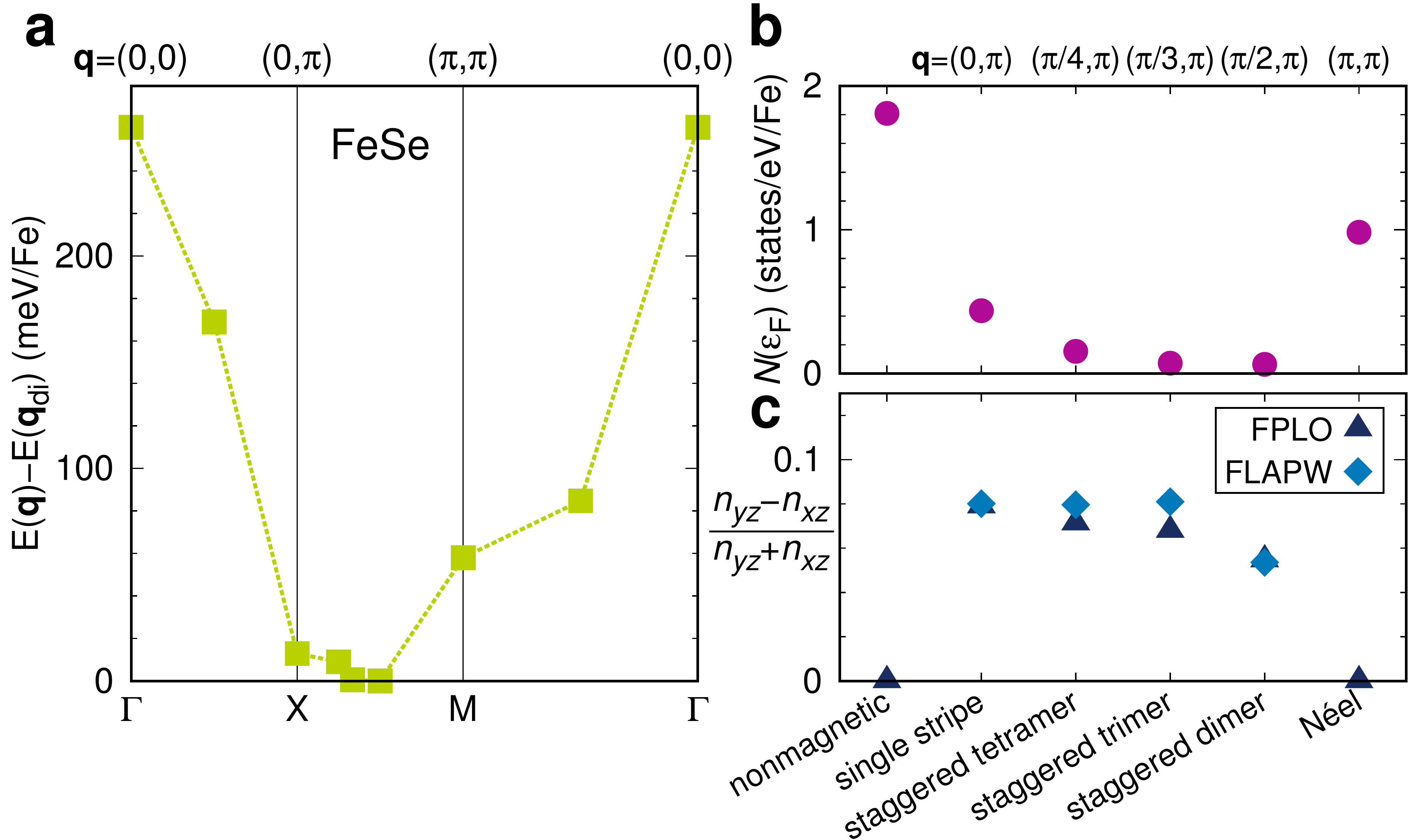}
  \caption{\textbf{Energies, density of states at $\varepsilon_F$ and
      orbital order of FeSe at 0 GPa pressure.} {\bf a} The energies
    of collinear magnetic configurations of FeSe at ambient pressure
    plotted as a function of $\mathbf{q}$. {\bf b} $N(\varepsilon_F)$
    for the lowest energy magnetic configurations compared to the
    nonmagnetic states indicates very small Fermi surfaces in
    fluctuating magnetic states. {\bf c} Ferro-orbital order in Fe
    $3d_{xz}/d_{yz}$ orbitals measured for several magnetic states.}
  \label{fig:fesl}
\end{figure*}

\newpage

\begin{table*}
  \centering
  \begin{tabular}{c c c c c} \toprule
    \multirow{2}{*}{Material} & $J_{1}$ & $J_{2}$ & $J_{3}$ & $K$ \\
    & \multicolumn{4}{c}{(meV)} \\ \midrule
    FeSe (0 GPa) & $123.1 \pm 6.5$ & $73.0 \pm 3.3$ & $18.3 \pm 1.8$ & $30.6 \pm 0.4$ \\
    FeSe (4 GPa) & $86.9 \pm 2.4$ & $51.9 \pm 1.2$ & $9.7 \pm 0.6$ & $15.7 \pm 0.2$ \\
    FeSe (9 GPa) & $51.1 \pm 0.7$ & $35.4 \pm 0.3$ & $4.9 \pm 0.2$ & $10.4 \pm 0.1$ \\
    FeTe & $50.7 \pm 3.6$ & $42.8 \pm 1.8$ & $24.4 \pm 1.0$ & $19.7 \pm 0.2$ \\ \bottomrule
  \end{tabular}
  \caption{\textbf{Heisenberg and biquadratic exchange parameters for 
      FeSe/Te.} The parameters for FeSe are reported at three different 
    pressures, 0 GPa, 4 GPa, and 9 GPa. FeTe is reported at 0 GPa. The 
    reported uncertainties indicate the inaccuracy of the fit to the {\Hb} model.}
  \label{tab:exchangeparameters}
\end{table*}

\begin{table*}
  \centering
  \begin{tabular}{c c c c c}
    \toprule
    Material & \textit{a} (\AA ) & \textit{c} (\AA ) & $z_{\text{Se}}$ & $z_{\text{Te}}$ \\ \midrule
    FeSe & 3.76976 & 5.52122 & 0.2688 &  \\
    FeTe & 3.81362 & 6.25381 &  & 0.2829 \\
    FeSe (4 GPa) & 3.6717 & 5.1943 & 0.2740 &  \\
    FeSe (9 GPa) & 3.6049 & 5.0304 & 0.2839 &  \\ \bottomrule
  \end{tabular}
  \caption{\textbf{Crystal parameters for FeSe and FeTe.} The structure 
    parameters and Wyckoff positions for FeSe/Te, with FeSe reported at
    three different pressures. The 4 and 9 GPa parameters are from 
    Ref.~\protect\onlinecite{Margadonna2009_PRB}, where we defined an effective tetragonal 
    lattice parameter $a^{*} = \protect\sqrt{ab}$ using the inplane 
    orthorhombic lattice parameters $a$ and $b$.}
  \label{tab:crystalstructure}
\end{table*}

\newpage

\begin{figure*}
  \centering
  \includegraphics[width=0.98\textwidth]{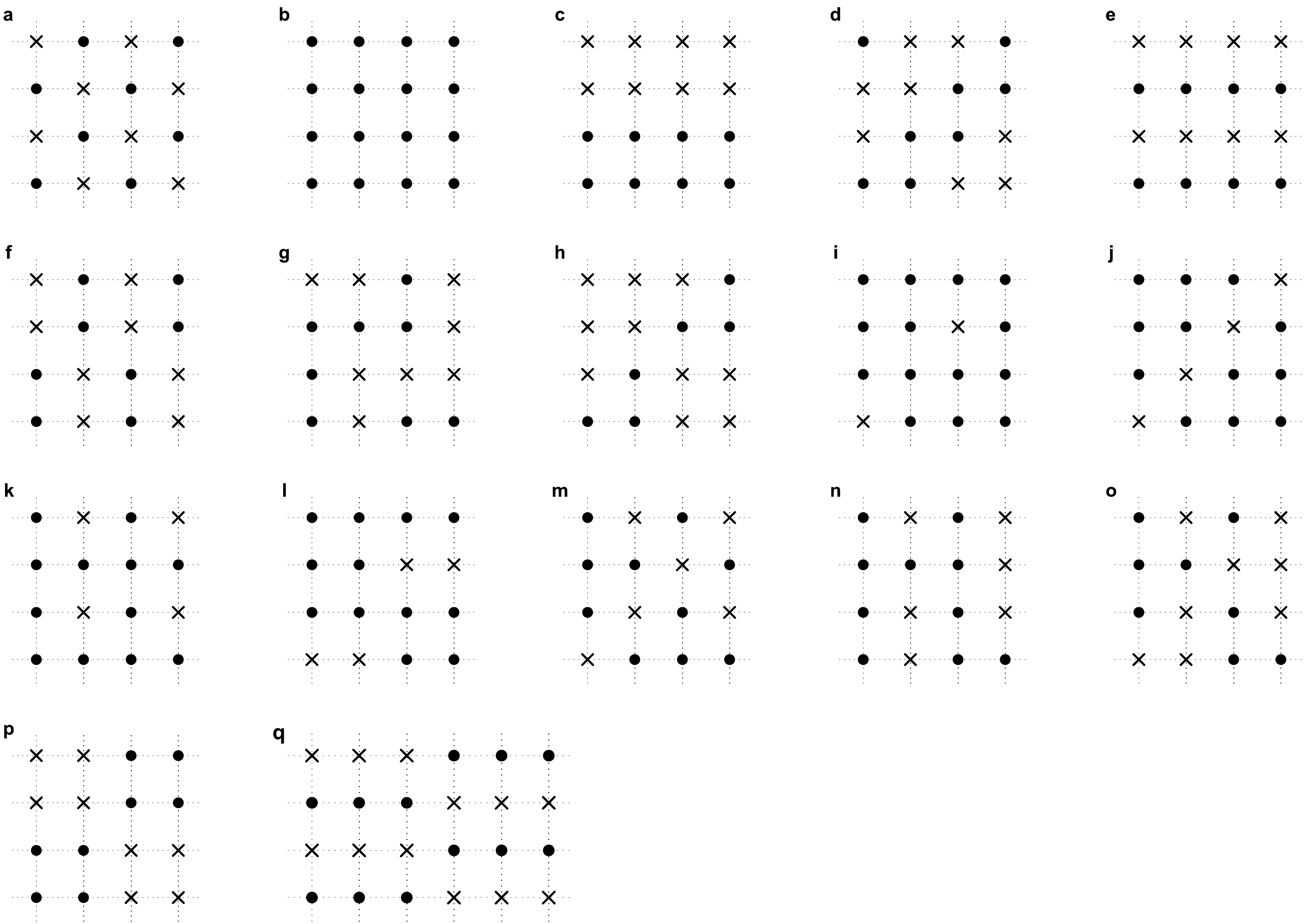}
  \caption{\textbf{Collinear magnetic structures} \textbf{a},
    Checkerboard (cb). \textbf{b}, Ferromagnetic (fm). \textbf{c},
    Parallel Stripes (parastr). \textbf{d}, Double Stripe
    (ds). \textbf{e}, Single Stripe (ss). \textbf{f}, Staggered
    Dimers (di). \textbf{g}, Zig-zag Stripes (zigzag). \textbf{h},
    dduuduuu. \textbf{i}, duuuuuuu. \textbf{j}, dduuuuuu. \textbf{k},
    udduuuuu. \textbf{l}, duuuduuu. \textbf{m}, ddduuuuu. \textbf{n},
    udduduuu. \textbf{o}, ddduduuu.  \textbf{p},
    plaquette. \textbf{q}, Staggered Trimers (tri).}
  \label{fig:suppmagstructures}
\end{figure*}

\begin{figure*}
  \centering
  \includegraphics[width=0.80\textwidth]{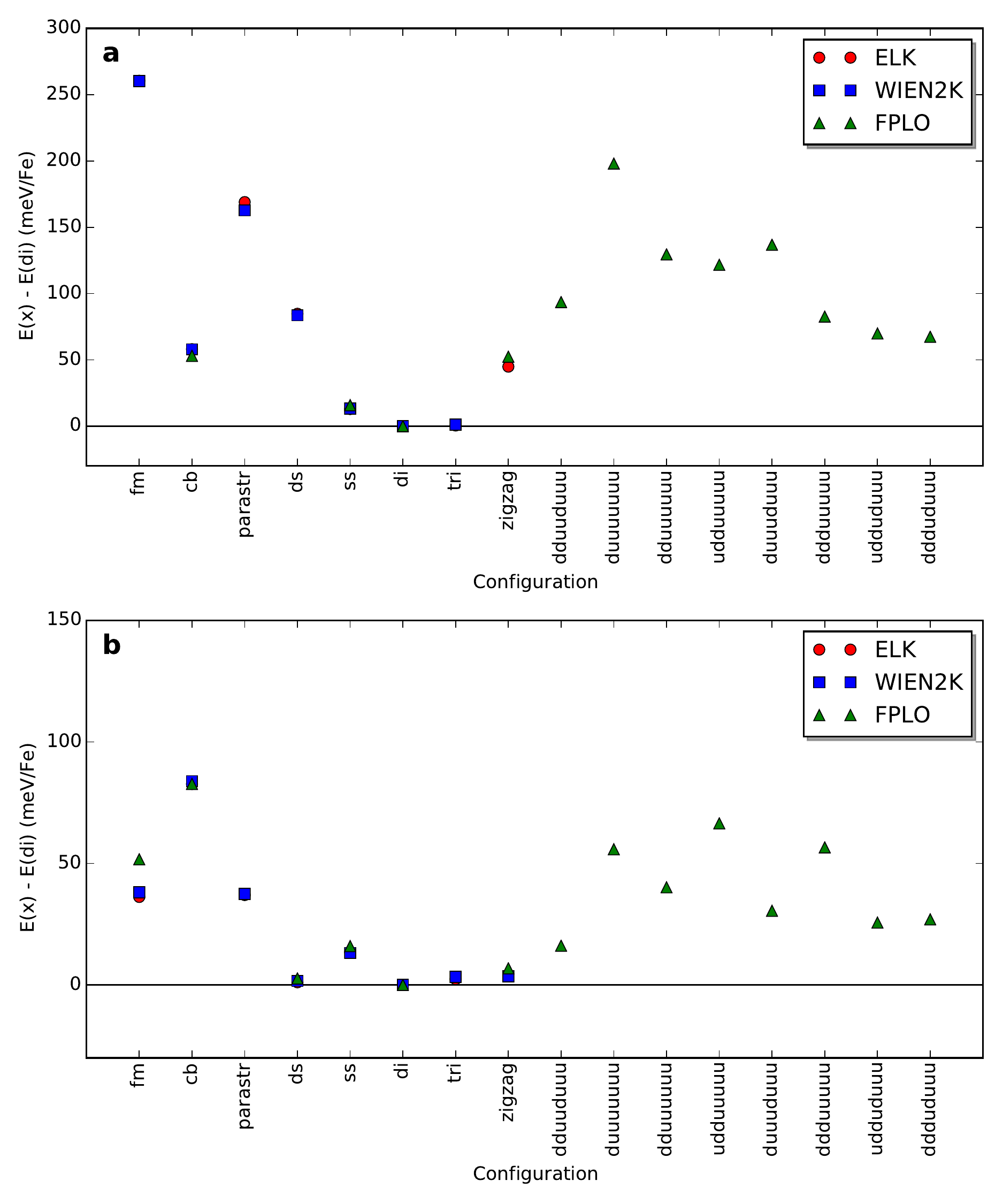}
  \caption{\textbf{Comparison of DFT energies} The DFT energies were
    calculated using \textsc{elk}, \textsc{wien2k}, and
    \textsc{fplo}. \textbf{a}, FeSe. \textbf{b}, FeTe. See
    Fig.~\ref{fig:suppmagstructures} for the different structures.}
  \label{fig:magenergies}
\end{figure*}

\begin{figure*}
  \centering
  \includegraphics[width=0.98\textwidth]{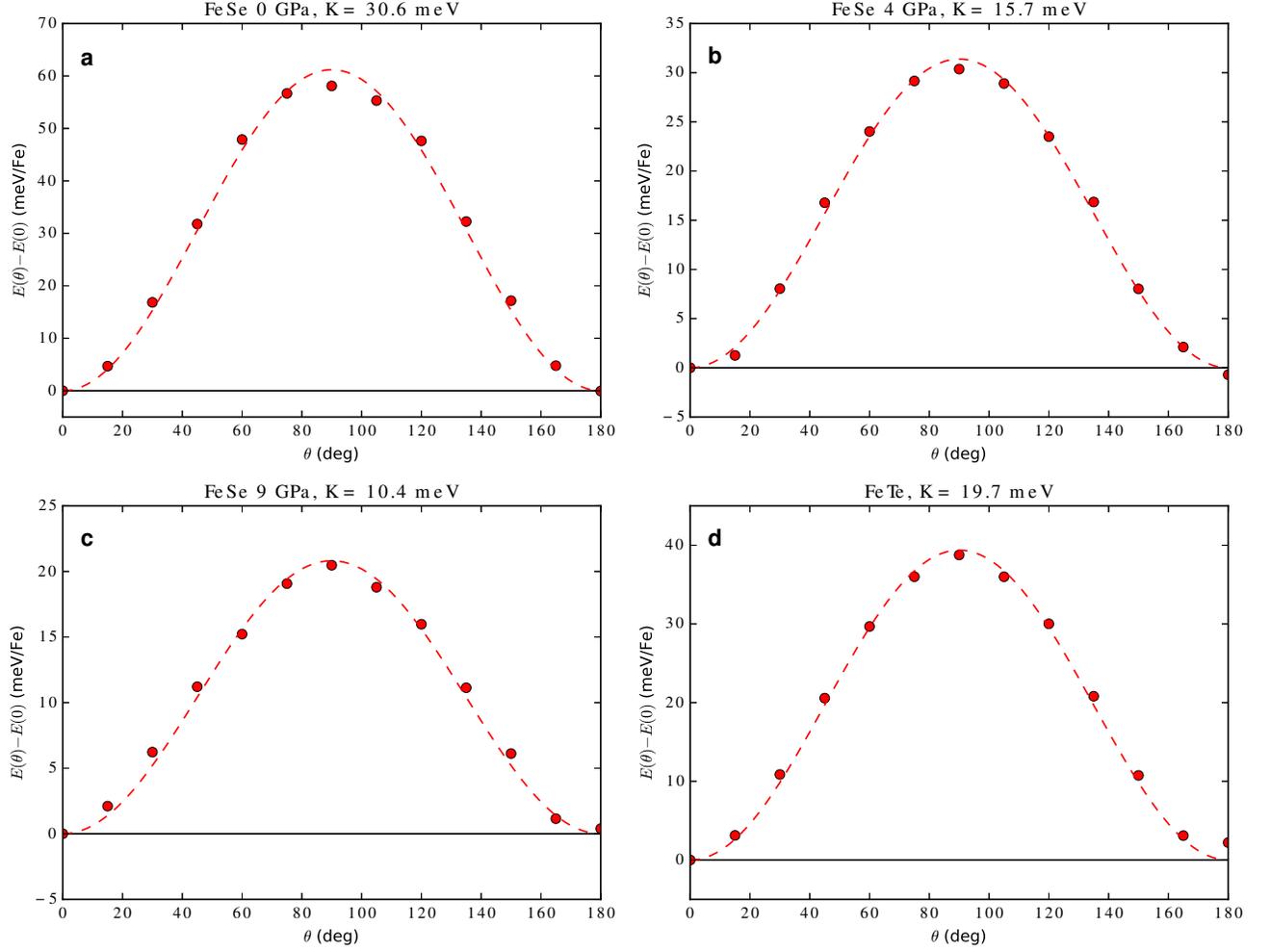}
  \caption{\textbf{Energies of noncollinear structures for FeSe and FeTe as a
    function of the rotation angle.} The dashed lines are the model fits. \textbf{a},
    FeSe at 0 GPa pressure. \textbf{b}, FeSe at 4 GPa pressure. \textbf{c}, FeSe at 9
    GPa pressure. \textbf{d}, FeTe.}
  \label{fig:noncollinear_energy}
\end{figure*}

\newpage

\begin{table*}
  \centering
  \begin{tabular}{ c c c c c } \toprule
    Configuration & $J_1$ & $J_2$ & $J_3$ & Const. \\ \midrule
    Checkerboard & -2 & 2 & 2 & 1 \\
    Single Stripes & 0 & -2 & 2 & 1 \\
    Double Stripes & 0 & 0 & -2 & 1 \\
    Dimers & -1 & 0 & 0 & 1 \\
    Trimers & -2/3 & -2/3 & 2/3 & 1 \\
    \bottomrule
  \end{tabular}
  \caption{\textbf{The $J_1$, $J_2$, and $J_3$ coefficients.} The coefficients 
    obtained by summing over neighbors of the magnetic structures used for fitting to the $J_1$-$J_2$-$J_3$-$K$ model.}
  \label{tab:heisenbergcoefficients}
\end{table*}

\begin{table*}
  \begin{tabular}{c c c c} \toprule
    Phase boundary & Ising & Heisenberg & $J_1$-$J_2$-$J_3$-$K$ \\ \midrule
    cb/di & $2J_{3} + 2J_{2} - J_{1}$ & & $2J_{3} + 2J_{2} - J_{1}$ \\
    cb/ss & & & \\
    ss/di & $2J_{3} - 2J_{2} + J_{1}$ & & $4J_{3} - 2J_{2} + J_{1} - 2K$ \\
    di/ds & $2J_{3} - J_{1}$ & & $2J_{3} - J_{1}$ \\
    cb/$\mathbf{q}=(\pi,Q)$ & & $4J_{3} + 2J_{2} - J_{1}$ & $4J_{3} + 2J_{2} - J_{1} - 2K$ \\
    ss/$\mathbf{q}=(\pi,Q)$ & & $4J_{3} - 2J_{2} + J_{1}$ & $4J_{3} - 2J_{2} + J_{1} - 2K$ \\
    cb/$\mathbf{q}=(Q,Q)$ & & $4J_{3} + 2J_{2} - J_{1}$ & $4J_{3} + 2J_{2} - J_{1} - 2K$ \\
    $\mathbf{q}=(\pi,Q)$/$\mathbf{q}=(Q,Q)$ & & $2J_{3} - J_{2}$ & $2J_{3} - J_{2} - K$ \\
    di/$\mathbf{q}=(\pi,Q)$ & & & $4K\left(2J_{3} - K\right) - \left(2J_{2} - J_{1}\right)^2$ \\
    di/$\mathbf{q}=(Q,Q)$ & & & $\left(J_{1} - J_{2} - 4J_{3} + 3K\right)^2 - J_{1} \left(J_{1} + J_{2} + K\right) + 2J_{1}^{2}$ \\
    ds/$\mathbf{q}=(\pi,Q)$ & & & $8J_{3} \left(2J_{3} - J_{1}\right) - \left(2J_{2} - J_{2}\right)^{2} - \left(2K - J_{1}\right)^{2} + J_{1}^{2}$ \\
    ds/$\mathbf{q}=(Q,Q)$ & & & $8J_{3}K + 4J_{2}K - 4K^{2} - J_{1}^{2}$ \\
    \bottomrule
  \end{tabular}
  \caption{\textbf{Analytical solutions for phase boundaries of the Ising, Heisenberg, and $J_1$-$J_2$-$J_3$-$K$ models.}}
  \label{tab:phaseboundaries}
\end{table*}

\end{document}